\documentclass[%
reprint,
superscriptaddress,
 amsmath,amssymb,
 aps,
pra,
]{revtex4-2}

\usepackage{soul}
\usepackage{graphicx}
\usepackage{dcolumn}
\usepackage{bm}
\usepackage{xcolor}
\graphicspath{{Figures/}}

\begin{document}

\preprint{}

\title{Reaching extreme fields in laser--electron beam collisions with XUV laser light}

\author{Brandon K. Russell}
\email{br2114@princeton.edu}
\affiliation{G\'{e}rard Mourou Center for Ultrafast Optical Science, University of Michigan, 2200 Bonisteel Boulevard, Ann Arbor, Michigan 48109, USA}
\affiliation{Department of Astrophysical Sciences, Princeton University, Princeton, New Jersey 08544, USA}

\author{Christopher P. Ridgers}
\affiliation{York Plasma Institute, School of Physics, Engineering and Technology,  University of York, York YO10 5DD, United Kingdom}

\author{Stepan S. Bulanov}
\affiliation{Lawrence Berkeley National Laboratory, Berkeley, California 94720, USA}

\author{Kyle G. Miller}
\affiliation{University of Rochester, Laboratory for Laser Energetics,
Rochester, New York 14623-1299 USA}

\author{Christopher Arran}
\affiliation{Physics Department, Lancaster University, Bailrigg, Lancaster, LA1 4YB, United Kingdom}
\affiliation{Cockroft Institute, Daresbury Laboratory, Keckwick Ln, Daresbury, Warrington, WA4 4AD, United Kingdom}
\affiliation{York Plasma Institute, School of Physics, Engineering and Technology,  University of York, York YO10 5DD, United Kingdom}

\author{Thomas G. Blackburn}
\affiliation{Department of Physics, University of Gothenburg, SE-41296 Gothenburg, Sweden}

\author{Sergei V. Bulanov}
\affiliation{ELI Beamlines Facility, The Extreme Light Infrastructure ERIC, Za Radnic\'{i} 835, Doln\'{i} B$\check{r}$e$\check{z}$any, 25241, Czech Republic}
\affiliation{National Institutes for Quantum and Radiological Science and Technology (QST),
Kansai Photon Science Institute, Kyoto, 619-0215 Japan
}

\author{Gabriele M. Grittani}
\affiliation{ELI Beamlines Facility, The Extreme Light Infrastructure ERIC, Za Radnic\'{i} 835, Doln\'{i} B$\check{r}$e$\check{z}$any, 25241, Czech Republic}

\author{John P. Palastro}
\affiliation{University of Rochester, Laboratory for Laser Energetics,
Rochester, New York 14623-1299 USA}

\author{Qian Qian}
\affiliation{G\'{e}rard Mourou Center for Ultrafast Optical Science, University of Michigan, 2200 Bonisteel Boulevard, Ann Arbor, Michigan 48109, USA}

\author{Alexander G. R. Thomas}
\affiliation{G\'{e}rard Mourou Center for Ultrafast Optical Science, University of Michigan, 2200 Bonisteel Boulevard, Ann Arbor, Michigan 48109, USA}

\date{\today}

\begin{abstract}

Plasma-based particle accelerators promise to extend the revolutionary work performed with conventional particle accelerators to studies with smaller footprints, lower costs, and higher energies. Here, we propose a new approach to access an unexplored regime of strong-field quantum electrodynamics by plasma wakefield acceleration of both charged particles and photons. Instead of using increasingly powerful accelerators and lasers, we show that photon acceleration of optical pulses into the extreme ultraviolet allows multi-GeV electrons to reach quantum nonlinearity parameters $\chi_e \gg 10$ with a high probability due to the reduced radiative losses. A significant fraction of photons produced in high-$\chi_e$ regions will propagate to detectors without generating pairs because of the reduction in the quantum rates. The photon spectra obtained may be used to characterize the predicted breakdown of strong-field quantum electrodynamics theory as it enters the fully non-perturbative regime.

 \end{abstract}

\maketitle

Quantum electrodynamics (QED) is probably the most successful quantum field theory ever developed, serving as one of the key components of the Standard Model of Particle Physics \cite{pdg.prd.2024}. However, particle and field interactions characterized by strong couplings have always presented a challenge for quantum field theories. QED is not an exception from this rule, as it was realized when charged particle and photon interactions with strong electromagnetic (EM) fields started to be considered \cite{ritus.jslr.1985}. Due to the nonlinearity of the interaction physics, standard perturbation theory does not apply and specific theoretical methodologies have had to be developed, which are usually referred to as strong-field quantum electrodynamics (SFQED) \cite{dipiazza.rmp.2012,Gonoskov_RevMod_2022,fedotov.pr.2023}. The processes described by SFQED are important in extreme astrophysics, for the next generation of lepton colliders, and high-intensity laser-matter interactions.  However, SFQED theory is itself conjectured to break down in the most extreme fields, and how to proceed theoretically is unknown as of now. The general failure of SFQED and prohibitively challenging theory within the fully non-perturbative regime represent a significant opportunity for experimental exploration and discovery in quantum physics.

The regime of SFQED occurs when the parameter $\chi_e = ||F^{\mu\nu} p_\nu||/mE_{cr}$ is large \cite{ritus.jslr.1985}, where $F^{\mu\nu}$ is the electromagnetic field tensor, $p_\nu$ is the electron 4-momentum, $E_{cr}=m^2 c^3/e\hbar\simeq 1.32\times10^{18}\,\text{V/m}$ is the QED critical field, $c$ is the speed of light, $e$ and $m$ are electron charge and mass, and $\hbar$ is the reduced Planck constant. The $\chi_e$ parameter describes the electric field strength relative to the QED critical field experienced by an electron in its rest frame.  Extreme fields in this context correspond to those for which the interaction can no longer be described by current SFQED theory. This regime has been conjectured \cite{narozhny.prd.1980} to appear when $\alpha\chi_e^{2/3}\sim 1$ (i.e., $\chi_e\approx 1600$), where $\alpha = 1/137$ is the fine structure constant. Studies of such extreme fields represent an intensity frontier of particle physics, in contrast to the usual high-energy frontier. Reaching this regime experimentally using conventional methods will require a significant technological leap. However, as we will argue in this paper, novel methods for the generation of extreme ultraviolet (XUV) light can greatly reduce the threshold for entering and studying the unexplored physics of this regime. The proposed method uses plasma-based acceleration which has seen recent widespread interest \cite{mirzaie.natphot.2024,Zhang_NatPhys_2025,Winkler_Nature_2025}.

Following pioneering experiments on the SLAC accelerator \cite{Burke_PRL_1997,Bula_PRL_1996}, there has been a recent surge of interest in SFQED studies \cite{fedotov.pr.2023}. The study of SFQED processes and the validation of models that have been proposed for these processes have formed a substantial part of the motivation for the construction of several multi-petawatt laser facilities around the world \cite{danson.hplse.2019,Zhang_PoP_2020,Gonoskov_RevMod_2022,mp3report.2022.arxiv,fedotov.pr.2023}, including ZEUS \cite{Nees_ZEUS}, Apollon \cite{papadopoulos.hpl.2016}, CoRELS \cite{yoon.optica.2021}, ELI NP \cite{gales.rpp.2018}, ELI-Beamlines \cite{weber.mre.2017}, and future NSF OPAL \cite{Bromage_HPLSE_2019}. While there exist several interaction schemes to study SFQED effects using lasers, the collision of a high-energy electron beam with a high-intensity laser pulse may be considered to be the most accessible. 
This is because plasma-based electron acceleration is a primary motivation for many laser facilities \cite{geddes.af6.2022,HEPAP23}, and this naturally provides a high-energy electron beam collocated with a PW-power laser.

Following recent experiments \cite{cole.prx.2018,poder.prx.2018,Mirzaie_NP_2024,los2025} approaching $\chi_e\sim1$, these facilities aim to reach a regime where $\chi_e>1$. In this regime, the stochastic emission of photons by electrons, usually referred to as the multi-photon Compton process \cite{nikishov.jetp.1964}, is the main cause of electron energy loss. The emitted photons upon interaction with the background laser field can decay into electron--positron pairs through the multi-photon Breit--Wheeler process \cite{reiss.jmp.1962,nikishov.jetp.1964}. Higher intensities may lead to EM shower cascades \cite{Gonoskov_RevMod_2022}, where the initial electron beam gives rise to multiple secondary photons, electrons, and positrons, effectively transferring its initial energy into the energy of secondary particles. Ultimately,  SFQED experiments will address the question of how  EM interactions behave under extreme conditions.

\section*{The breakdown of SFQED}

 The Ritus--Narozhny conjecture \cite{narozhny.prd.1980,ritus.jslr.1985} states that since semiclassical radiative corrections scale as $\alpha\chi_e^{2/3}$, at such values of $\chi_e$ the breakdown of the perturbative expansion used to describe multi-photon Compton and Breit--Wheeler processes should occur. It was recently found \cite{podszus.prd.2019,ilderton.prd.2019} that this behavior of radiative corrections is not universal and rather depends on whether the process is dominated by laser intensity or electron energy. The former results in the probability of the multi-photon Compton process to asymptotically approach $P\sim \alpha\chi_e^{2/3}$, while the latter results in $P\sim \alpha a_0^2\log(\chi_e)/\chi_e$. The logarithmic behavior is due to the breakdown of the locally constant crossed-field approximation (LCFA), which occurs when the number of absorbed photons from the background field is no longer much larger than one, $a_0^3/\chi_e\gg 1$. Here, the parameter $a_0$ is defined as $a_0 = eE_0/mc\omega_0$, $E_0$ is the electric field strength and $\omega_0$ is the laser frequency. This parameter can represent the coupling strength between the charged particles and classical laser field, or the magnitude of the electron quiver momentum relative to $mc$. 

\begin{figure}
    \includegraphics{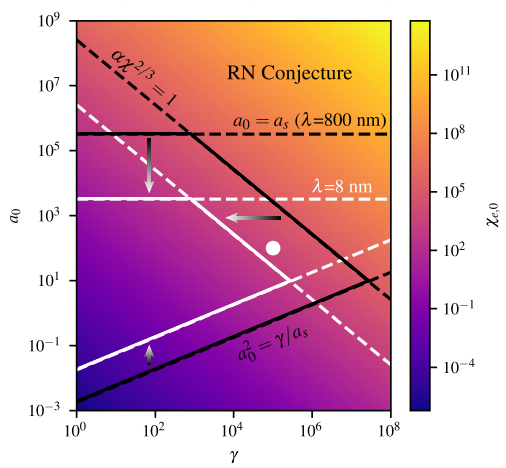}
    \caption{\textbf{Regimes of perturbative SFQED breakdown.} The dependence of $a_0$ on $\gamma$ satisfying either $\alpha\chi_e^{2/3}=1$ (RN) or $a_0^2=\gamma/a_S$. Curves are shown for $\lambda=800$ nm in black and $\lambda=8$ nm in white, i.e., $\Omega=100,\; \xi=1$. The white circle is an example point at $a_0=100$ and $\gamma = 10^5$. By increasing the laser frequency, different regimes of the breakdown of perturbative SFQED can be studied.  
    }
    \label{figure7}
\end{figure}

\begin{figure*}
    \centering
    \includegraphics{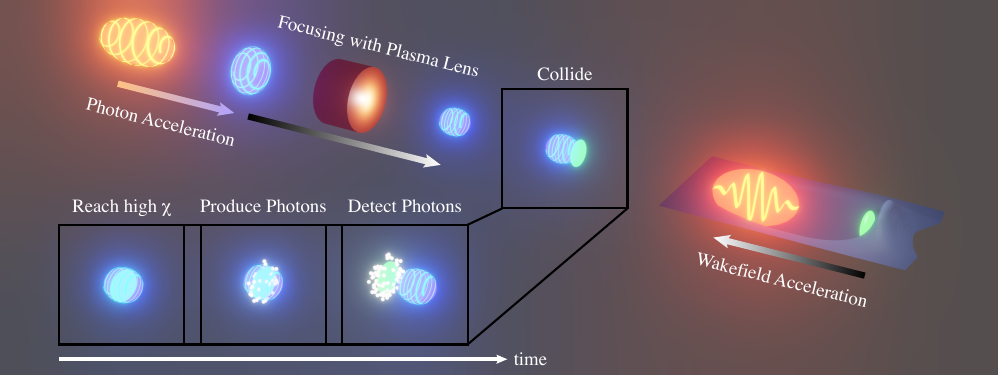}
    \caption{\textbf{An experimental concept for reaching the fully perturbative regime of SFQED.} A laser pulse is injected into the wake of a particle or laser beam propagating through an underdense plasma. In the wake, the pulse experiences a temporal density gradient and therefore a refractive index gradient that allows the pulse to gain energy and the wavelength to decrease from the original wavelength up to XUV wavelengths through the process of ``photon acceleration'' \cite{Wilks_PRL_1989,Murphy_PoP_2006}. The accelerated pulse is focused to the collision point using a plasma lens \cite{Palastro_PoP_2015,Li_PRR_2024}. This is necessary to maximize the pulse intensity in the collision; a plasma lens is capable of focusing the already intense laser pulse. Simultaneously an electron beam is accelerated through laser wakefield acceleration (or using a conventional accelerator) to multi-GeV energies. The accelerated electron beam collides head on with the XUV pulse, generating an extremely high-$\chi_e$ interaction. 
    }
    \label{fig:exp_concept}
\end{figure*}

In a head-on collision with a plane wave laser, $\chi_e\approx 2\gamma a_0 \hbar\omega_0/mc^2$, where $\gamma$ is the electron Lorentz factor.  So for a given laser frequency $\omega_0$ it is possible to define a critical vector potential of the QED critical field, $a_S=m c^2/\hbar \omega_0$, such that $\chi_e \approx 2\gamma a_0/a_S$. In order to visualize these regimes of extreme interactions (i.e., $\alpha\chi_e^{2/3}=1$ and  $a_0^2= \gamma/a_S$), Fig. \ref{figure7} shows the corresponding curves in the $(\gamma,a_0)$ plane. The parameter space usually considered in  SFQED studies is limited from above by $\alpha\chi_e^{2/3}=1$ and $a_0=a_S$ (which denotes the laser electric field strength reaching the QED critical value) and by the $a_0^2= \gamma/a_S$ curve from below. The $\alpha\chi_e^{2/3}=1$ and  $a_0^2= \gamma/a_S$ curves intersect at $\gamma=2^{-2/3}\alpha^{-1}a_S$ and $a_0=2^{-1/3}\alpha^{1/2}$, where the value of $\gamma$ depends only on laser frequency and $a_0$ is a constant. This value of $\gamma$ can be interpreted as $\lambdabar/\gamma\sim r_e$, i.e., the reduced laser wavelength in the electron rest frame is of  order  the classical electron radius, which indicates both the breakdown of the SFQED semi-classical perturbation theory and the LCFA. What happens near the intersection of these scalings is unknown, and it is crucial to explore these limits with experiments to understand the breakdown of the SFQED semi-classical perturbation theory. 

From an experimental perspective, reaching the required $\chi_e\sim 1600$ seems like a  distant goal with current laser facilities. Naively, the next generation of multi-PW class laser facilities might be expected to reach this regime if electron beams in excess of 100 GeV were available for the collision \footnote{This regime may also be reached in the collision of 100 GeV electron beams \cite{Yakimenko_PRL_2019}, though a very challenging focusing of the beams is assumed.}, which by itself is a grand challenge for laser--plasma based accelerators \cite{geddes.af6.2022,HEPAP23}. Even in this case, however,  electrons will strongly radiate in the laser fields prior to the peak intensity, leading to, on average, a substantially lower value of $\chi_e$ than would be predicted from the peak field and maximum electron energy \cite{Blackburn_NJP_2019,jirka.pra.2021}. This limitation is severe; simply increasing $a_0$ for the pulse will not raise the achieved $\chi_e$ for an average electron---since $\chi_e\sim a_0$ but the radiated energy scales as $a_0^2$---so the average electron will simply radiate its energy away faster. However, photon emission is stochastic for large $\chi_e$; therefore, a small fraction of the electron beam population may make it to the peak field without radiating---an effect known as `straggling' \cite{blackburn.prl.2014,harvey.prl.2017}. Both the radiated energy and the straggling effects are strongly dependent on the laser wavelength, and thus the utilization of laser light with non-optical frequencies might make the regime of extreme interaction not only accessible but also observable in future experiments \cite{Baumann_SR_2019,Zaim_PRL_2024}.

In SFQED studies, several methods of electromagnetic field intensification have been proposed (see Ref. \cite{Gonoskov_RevMod_2022} and references cited therein). Some of them lead to a significant frequency upshift of the generated radiation, pushing it into the XUV and hard X-ray regimes (see e.g. Refs. \cite{bulanov.ufn.2013,marklund.hplse.2023}). In particular, it was recently shown that by using a driver electron beam propagating through a tailored plasma density profile and inciting a plasma wave, the photon distribution of a laser pulse placed at a particular position in the trailing wake can be upshifted in frequency over large distances \cite{Sandberg_PRL_2023} (see Fig. \ref{fig:exp_concept}). In simulations, frequency shifts were demonstrated to XUV photon energies while approximately maintaining the $a_0$ of the pulse. Notably, this method also preserves the number of cycles in the laser pulse and the polarization, making comparison with theory more transparent. As it will be shown, this photon acceleration can greatly increase $\chi_e$ in an electron--laser collision.

In what follows, we evaluate what can be achieved in the study of extreme SFQED effects by combining  two different types of plasma--based acceleration: that of electrons and that of photons by either a laser or a particle beam. The basic principle of the proposed setup is outlined in Fig. \ref{fig:exp_concept}. The optical pulse is accelerated up to XUV light in the wakefield of an electron beam. The XUV pulse is focused by a plasma lens and then collides with a counter-propagating multi-GeV electron beam that is accelerated by the laser wakefield. Such a configuration can be achieved at future multi-PW laser facilities with electrons reaching $\chi_e$ values characteristic of the extreme interaction conditions. We derive conditions necessary for the electrons to reach the peak $\chi_e$ and emit high-energy photons, which can be used to diagnose high-$\chi_e$ effects. We additionally consider whether these photons would make it to a detector, or if they generate pairs and therefore cannot be used as a high-$\chi_e$ diagnostic.
\newline

\section*{Results}

\subsection*{The Utility of Photon Acceleration} 

The maximum achievable value of $\chi_e$ in the collision of an electron beam with a laser pulse occurs when the peak normalized vector potential $a_0$ of a pulse with central frequency $\omega_0$ interacts with the most energetic electrons in the beam with Lorentz factor $\gamma_0$. This is denoted by $\chi_{e,0}$ and is,

\begin{equation}
\chi_{e,0} = \frac{2a_0\gamma_0\hbar\omega_0}{mc^2}.
\end{equation}
When an optical pulse is converted to an XUV pulse, the frequency increases by a factor $\Omega = \omega_X/\omega_O$ while $a_0$ changes by a factor $\xi = a_X/a_O$, where subscripts X and O are used to refer to XUV and optical pulses, respectively. We will denote the maximum $\chi_e$ achievable in the collision with the optical pulse by $\chi_e^{O}$  and the XUV pulse by $\chi_e^{X}$ with

\begin{equation}
\chi_e^{X} = \frac{2a_X\gamma_0\hbar\omega_X}{mc^2} = \frac{2a_O\gamma_0\Omega\hbar\omega_O\xi}{mc^2} = \Omega\xi\chi_e^{O}.
\end{equation}
In the case of photon acceleration, we expect $\Omega>\mathcal{O}(10)$ and $\xi\sim1$ \cite{Sandberg_PRL_2023} so $\chi_e^{X}>\chi_e^{O}$.
\newline

Let us show how photon acceleration makes reaching the extreme regimes of SFQED easier in terms of the laser energy needed. If we start from the interaction of a 50-GeV electron beam ($\gamma_0=10^5$) with a $a_0=100$ laser pulse, then one can boost this interaction to the $\alpha\chi_e^{2/3}=1$ regime by either increasing the electron beam energy, the laser intensity, or both. If we assume that the electron beam is accelerated via laser--plasma acceleration (LPA) \cite{esarey.rmp.2009}, then its energy scales as laser pulse energy to the power of two thirds \cite{turner.epjd.2022,bulanov.arxiv.2023}. Thus, the laser energy needed to reach the $\alpha\chi_e^{2/3}=\alpha(2\gamma'a_0'/a_S)^{2/3}=1$ regime would be $U'=U_e(\gamma'/\gamma_0)^{3/2}+U_a(a_0'/a_0)^2$, where $U_e$ is the laser energy needed to accelerate electrons to $\gamma_0=10^5$ and $U_a$ is the laser energy needed to achieve $a_0=100$ at focus. For $U_e=300$ J and $U_a=100$ J the minimum energy needed is $\sim 6$ kJ (which is realized by boosting $a_0$ to $a_0'=507$ and $\gamma_0$ to $\gamma'=5\times 10^5$). This is a 15 times energy increase from the 400 J used to generate the initial interaction. As shown in Fig.~\ref{figure7} photon acceleration shifts the $\alpha\chi_e^{2/3}=1$ and $a_0^2= \gamma/a_S$ curves to the left, so that the $\gamma_0=10^5$, $a_0=100$ interaction is well into the $\alpha\chi_e^{2/3}>1$ regime for $\Omega=100$ and $\xi=1$. From the point of view of laser energy required for photon acceleration, another at least 50-GeV electron beam is needed \cite{Sandberg_PRL_2023}, which translates into a factor of two in laser energy increase. The energy increase is significantly lower than the factor of 15 needed for boosting the electron beam energy and laser intensity. 

We also note that for the interaction to reach the $a_0^2= \gamma/a_S$ regime with $\lambda=800$ nm and $a_0>1$, the electron energy needed should exceed 150 GeV. The use of photon acceleration relaxes this number, e.g., for $\Omega=100$, the electron energy needed is 1.5 GeV. Figure \ref{fig:exp_concept} shows a concept for how photon acceleration can be used to reach the high-$\chi_e$ regime.

\begin{figure}
    \centering
    \includegraphics{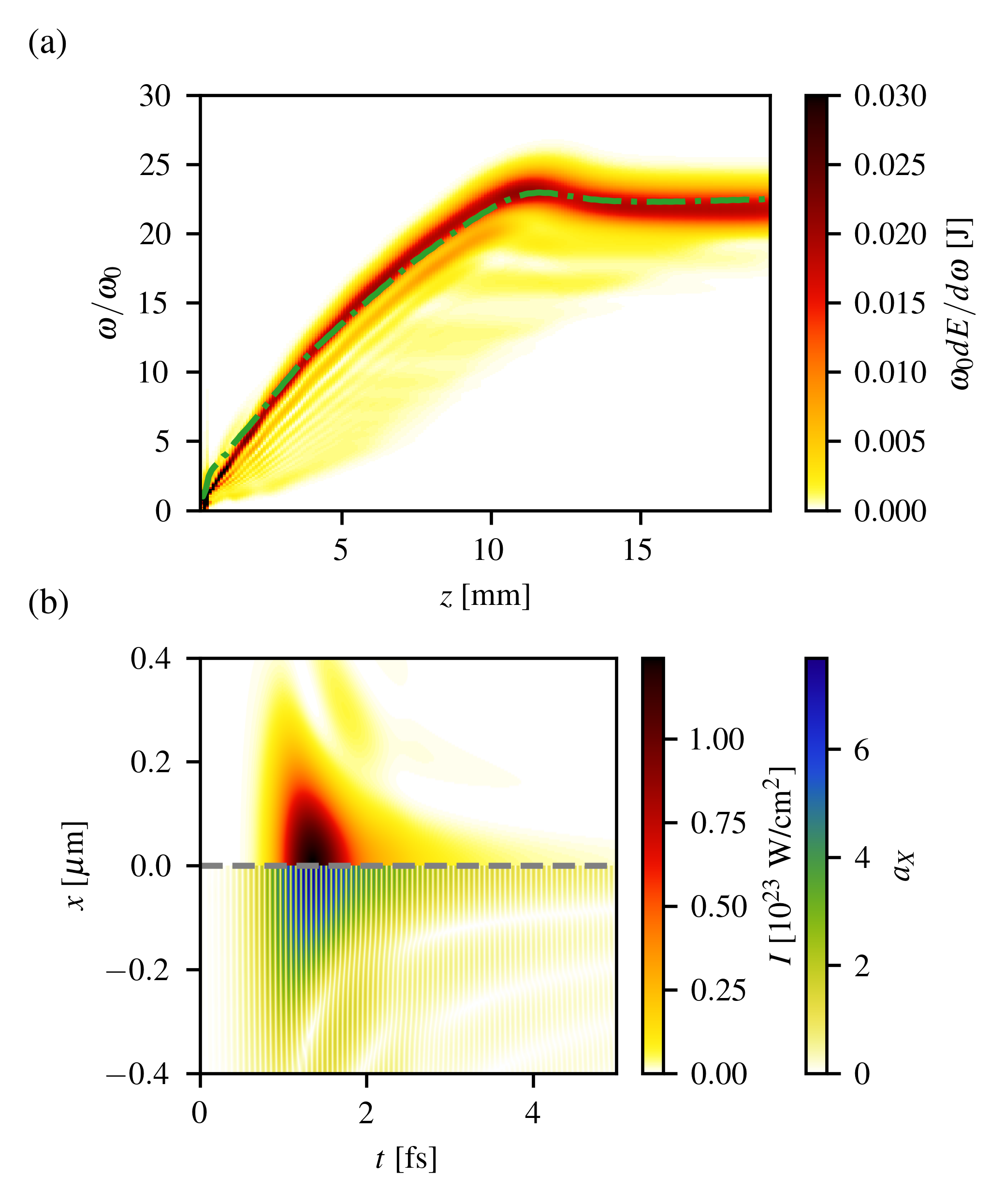}
    \caption{ \textbf{Determining realistic laser properties from photon acceleration.} Quasi-3D, boosted-frame \textsc{Osiris} simulation demonstrating the acceleration of an initially $a_0 = 2.8$, $\lambda_L = 800$ nm pulse in the wake of a 50-GeV electron beam propagating through a tailored density plasma. (a) The spectrum of the pulse, where the dashed–dotted line is the central frequency.  (b) The electric field (lower) and intensity (upper) of the pulse after focusing by a plasma lens. 
    }
    \label{fig:photon_accel}
\end{figure}

\begin{figure*}
    \centering
    \includegraphics{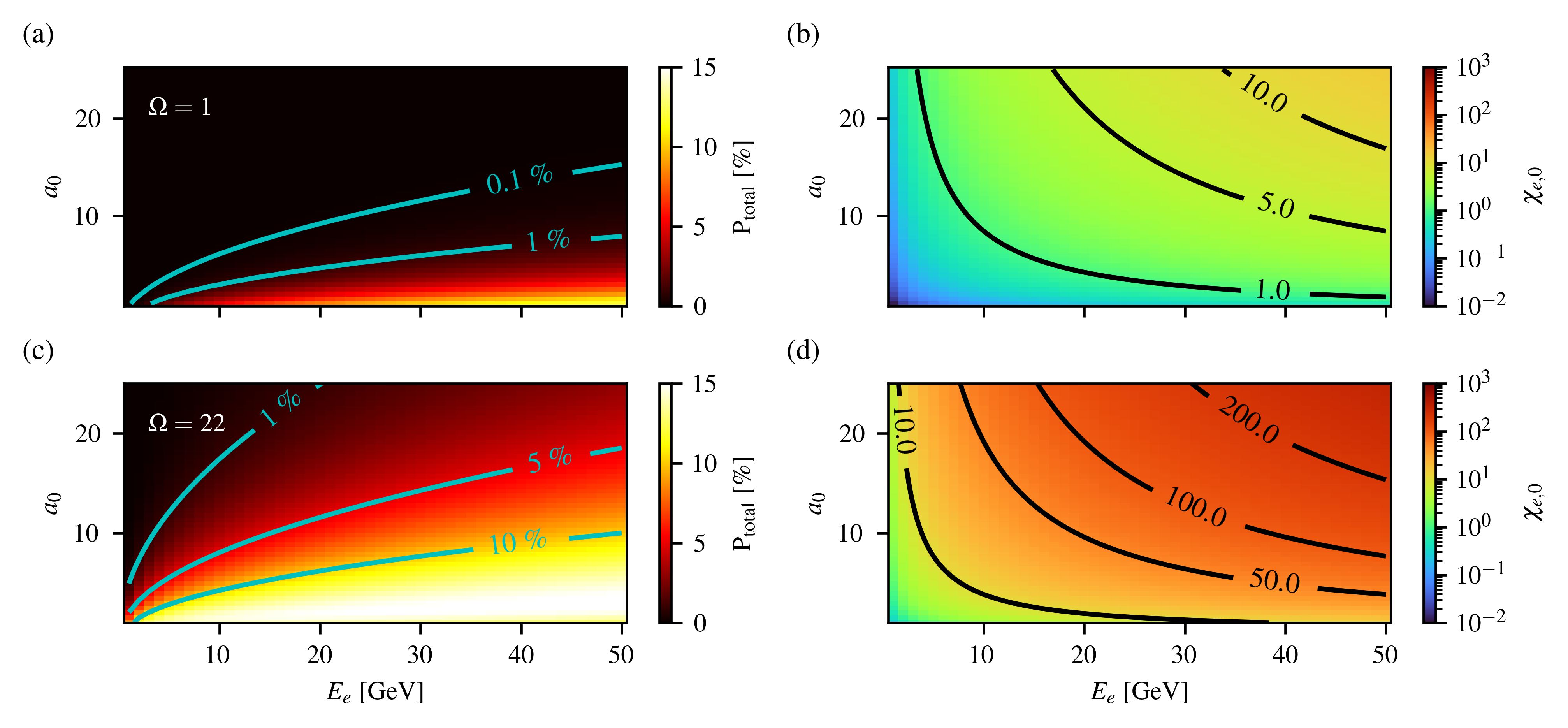}
    \caption{\textbf{Comparison of optical and XUV pulse probabilities and achievable $\chi_e$.} Total probability of electrons propagating to high-$\chi$ in a laser pulse, emitting photons, and these photons propagating out of the pulse without generating pairs for a frequency upshift factor of (a–b) 1 and (c–d) 22. The number of laser cycles $N = \omega\tau_L/2\pi = 11.625$ where $\tau_L$ is the full-width half-maximum duration.
    }
    \label{fig:P_total}
\end{figure*}

\subsection*{Probabilities for Measuring High-$\chi_e$ Effects} 

As we will show in this section, not only does XUV laser light reduce the energy requirements for reaching high $\chi_e$, but it also greatly increases the probability of observing high-$\chi_e$ effects. There are three primary conditions that must be met to reach and measure high-$\chi_e$ effects. First, the electrons cannot radiate a significant part of their energy before reaching the peak of the pulse. Second, these electrons must radiate near the peak of the pulse to generate a measurable signature of the high-$\chi_e$ effects. Finally, the photons must make it out of the pulse and to a detector without generating pairs. In the following, we will consider the probabilities of these processes through theory and simulations under realistic conditions for photon-accelerated pulses.

To first establish what $\Omega$ and $\xi$ can be reasonably be expected, simulations were performed using the \textsc{Osiris} 4.0 particle-in-cell (PIC) code leveraging the quasi-3D and boosted-frame capabilities. Starting with an $a_0 = 2.8$, $\lambda_L = 800$ nm central wavelength pulse in the wake of a 50-GeV electron beam, the frequency upshifts by $\Omega = 22$ and, through focusing using a plasma lens, $a_0 = 8$ ($\xi \approx 2.9$). These results are shown in Fig.~\ref{fig:photon_accel} (see Methods for additional details). These parameters will be used in the remainder of the paper. 

The total probability is the product of the three probabilities, which are found by integrating the respective rate equations (see Methods for details). Figure~\ref{fig:P_total} shows the total probability for a range of $a_0$ and $\gamma_0$ and for (a-b) $\Omega = 1$ and (c-d) 22. Within the plotted parameter space, neither the optical nor XUV pulse reaches $\chi_e = 1600$. The XUV pulse reaches $\chi_e > 100$, however, as we argue later this is sufficient to test the breakdown of perturbative SFQED. Furthermore, we observe a significant $P_{total}\sim 10\%$ for the XUV pulse. The biggest detractor from the total probability is the probability that electrons do not emit as they propagate to the peak of the laser pulse. This probability is maximized at a particular $\chi_e$ by maximizing the electron beam energy and minimizing $a_0$. However, the Ritus--Narozhny conjecture is only valid in the limit of $a_0\gg 1$, thereby limiting the maximum beam energy and pulse frequency.

\subsection*{Simulated Probabilities}

\begin{figure}
    \centering
    \includegraphics{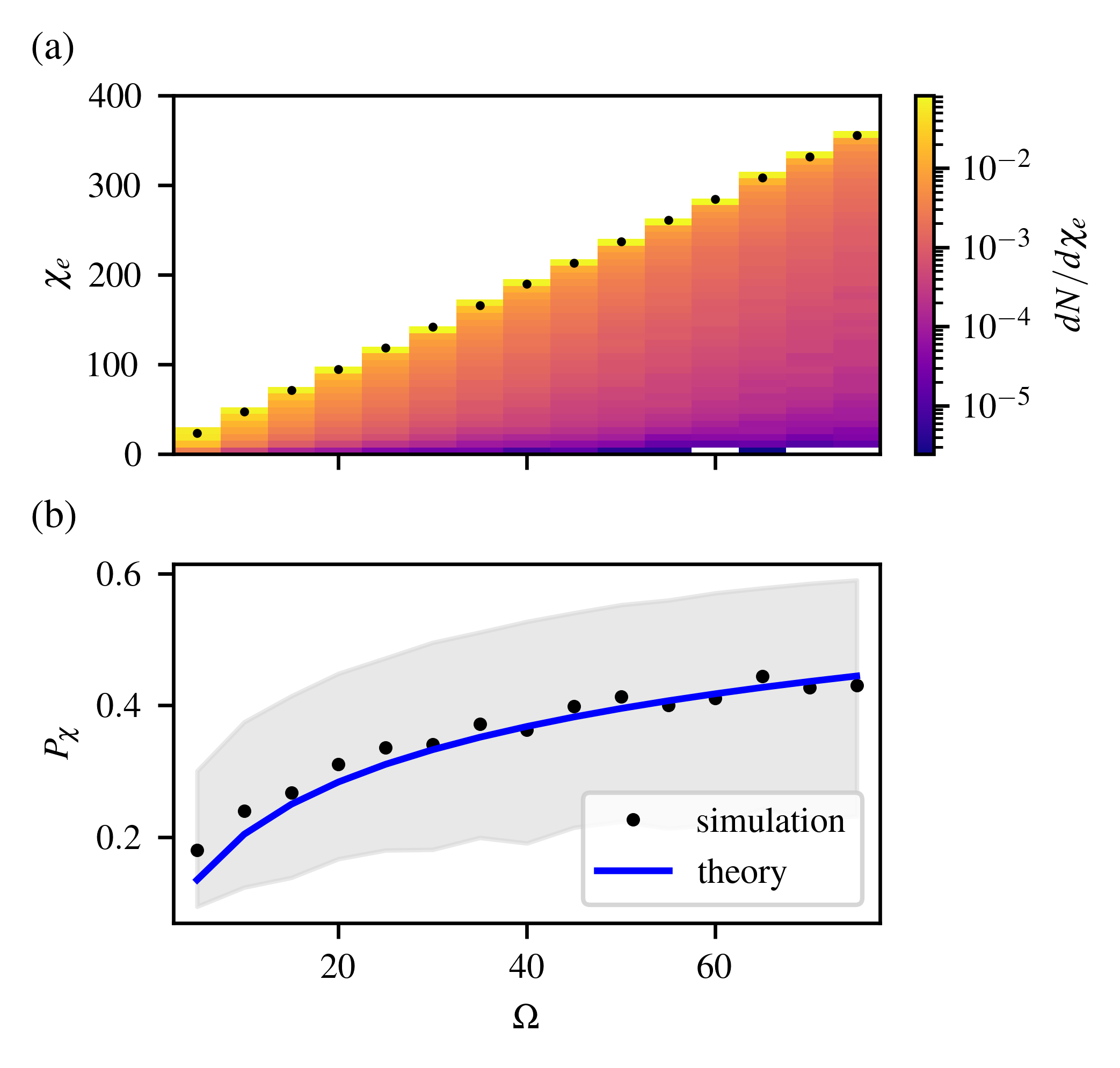}
    \caption{\textbf{Fraction of electrons reaching the peak $\chi_e$ as function of the frequency upshift $\Omega$.} Distributions of maximum $\chi_e$ reached by 50-GeV electrons in collision with an $a_0 = 8$, $N = 11.625$ circularly polarized laser pulse with varying frequency ($\Omega$). Black dots in (a) show $\chi_{e,0}$ for each value of $\Omega$. The fraction of particles within 0.09\% of $\chi_{e,0}$ is shown in (b) as black dots with a gray filled region showing the fraction if 0.05\% and 1\% are chosen instead. This is plotted alongside the probability calculated from Eq.~\eqref{Eqn:P_ne}.
    }
    \label{fig:P_epoch}
\end{figure}

\begin{figure*}
    \centering
    \includegraphics{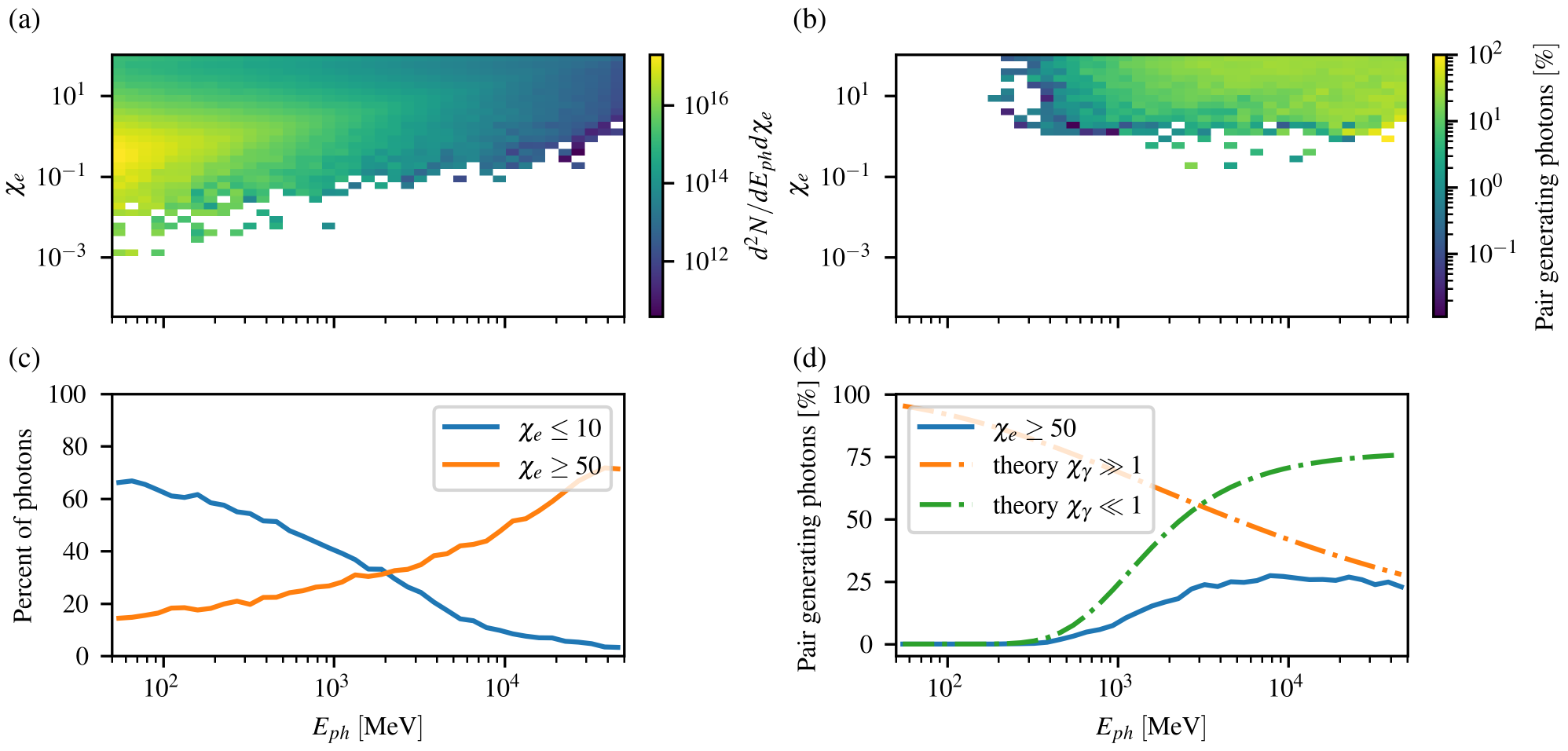}
    \caption{\textbf{The energy of photons emitted by high-$\chi_e$ electrons and the photons lost to pair creation.} (a) Distribution of photons as a function of energy and the $\chi_e$ at which they were produced. Photons were produced in the collision of a 50-GeV electron beam with an $a_0 = 8$, $N=11.625$ circularly polarized laser pulse with $\Omega = 22$. (c) The percentage of photons with a particular energy produced in regions with $\chi_e\leq 10$ and $\chi_e \geq 50$. (b) Percent of photons that generated pairs as a function of energy and the $\chi_e$ at which the photons were produced. (d) The percentage of photons that were produced in regions of $\chi_e \geq 50$ that generated pairs. This is compared with the theory for generating pairs calculated in the limit of $\chi_\gamma \gg 1$ (Eq.~\eqref{Eq:tau_bw}) and $\chi_\gamma \ll 1$ (Eq.~\eqref{Eq:BW_low_chi}).  
    }
    \label{fig:EPOCH_photon_chi}
\end{figure*}

The details of the collision are further understood through 1D PIC simulations in the code \textsc{Epoch} (see Methods for additional details). The code was modified to track the peak $\chi_e$ of all electrons in the collision, the $\chi_e$ at which photons are generated, and which photons generate pairs. By scanning $\Omega$, we find that as predicted by the theory, both $\chi_e$ and the fraction of the distribution that reaches the peak $\chi_e$ increase with $\Omega$ (see Fig.~\ref{fig:P_epoch}). Figure~\ref{fig:EPOCH_photon_chi}(a,c) shows that the high-$\chi_e$ electrons dominate the emission of the highest-energy photons. However, even for 100-MeV photons, almost 20\% come from regions of $\chi_e\geq 50$. It will be a significant challenge to measure such high-energy photons, and novel detectors and analysis techniques will need to be developed to do so. The high-energy photons also dominate the generation of pairs: as shown in Fig.~\ref{fig:EPOCH_photon_chi}(b,d), approximately 25\% of 10-GeV photons are lost. However, taken together, Figs. \ref{fig:P_epoch} and \ref{fig:EPOCH_photon_chi} demonstrate that photon acceleration allows for a large fraction of the initial beam to reach high-$\chi_e$, generate a significant amount of photons---especially at the highest energies---and for these photons to propagate to a detector.

\section*{Discussion}

Although we showed that increasing laser frequency greatly increases the achievable $\chi_e$ and the number of particles that reach $\chi_{e,0}$, these results were obtained using a number of approximations and assumptions, which are discussed below.

First, the RN conjecture \cite{ritus.jslr.1985,narozhny.prd.1980} states that the radiative corrections in SFQED scale as $\alpha\chi_e^{2/3}$, making the semiclassical perturbation theory invalid at $\chi_e=1600$ ($\alpha\chi_e^{2/3}=1$). However, this means that these radiative corrections should be taken into account at significantly lower values of $\chi_e$, since at $\chi_e\approx 50$ their contribution is already at $\sim 10\%$ \cite{mp3report.2022.arxiv}. Thus, it is not correct to expect Fig. 4 at high $\chi_e$ to be accurate, because it was calculated using only the lowest-order approximation.   

Second, there is no universal high-$\chi_e$ behavior, as it was shown in Refs. \cite{podszus.prd.2019,ilderton.prd.2019}. For the probability of the Compton process as $\chi_e\to\infty$ at fixed frequency $\omega_0$, $P\sim \alpha\chi_e^{2/3}$ for $a_0\rightarrow \infty$, and $P\sim \alpha a_0^2\log(\chi_e)/\chi_e$ for $\gamma\rightarrow \infty$. However, PIC codes are able to capture only the $\alpha\chi_e^{2/3}$ behavior due to the use of the LCFA to model the SFQED processes. Additionally, note that these limits are characterized by completely different scalings of the Compton process probability with the EM field frequency. While $\alpha\chi_e^{2/3}$ scales as $(\xi\Omega)^{2/3}$, the $a_0^2\log(\chi_e)/\chi_e$ scales as $\xi\log(\Omega\xi)/\Omega$. Whereas the first case gives increase with the parameters of photon acceleration considered in this paper, the second one leads to the decrease of the probability of the Compton process with photon acceleration. Thus, these two limits can be distinguished with the help of the photon acceleration. 

\section*{Conclusions}

In conclusion, we used the process of plasma-wakefield-driven photon acceleration to obtain an XUV laser pulse, which can be used to probe the extreme regimes of SFQED. As is stated in the literature, it is challenging for the electrons to reach maximum possible values of $\chi_e$ during the interaction with a high-intensity laser pulse due to the fast depletion of electron energy from photon emission. We showed that frequency upshifted laser pulses allow a significant portion of the colliding electrons to reach the peak of the pulse without losing any energy.

The use of frequency upshifted laser pulses opens the possibility to study the extreme regimes of SFQED by observing photons emitted by these electrons at the peak of the XUV laser pulse. This also requires that those photons do not convert into electron-positron pairs while propagating in the XUV laser pulse. Thus, the signal that we propose to observe is actually a product of three probabilities: (1) for the electron to reach the peak of the XUV laser pulse without emitting a photons, (2) for the electron to emit a photon at the peak of the pulse, (3) for this photon to reach the detector without converting into an electron-positron pair. It turns out that the probability of observing such photons in not negligible. Approximately the same number of photons as the number of initial electrons in the colliding beam is able to reach the detector 

Whereas one of the main goals of such a study would be the detection of the radiative corrections to the Compton process and confirmation of its scaling with laser intensity and particle energy as $\alpha\chi_e^{2/3}$ (RN conjecture), the photon acceleration opens up one more exciting possibility for SFQED studies. It allows to probe the breakdown of the locally constant field approximation ($a_0^2\gg \gamma/a_S$) in the extreme regime where $a_0>1$, and to possibly observe the logarithmic behavior of the multi-photon Compton process probability. The methods outlined in this paper give a clear direction for the near-term development of laser facilities to study these completely unexplored regimes of quantum physics.

\section{Methods}

\subsection{Probabilities}

Here we outline the equations that were used to generate Fig.~\ref{fig:P_total} and were also used to compare with simulations in Figs.~\ref{fig:P_epoch} and \ref{fig:EPOCH_photon_chi}.

\subsubsection{Probability of non-emission}

The fraction of the distribution that does not emit before the peak of the laser pulse is given by, 
\begin{equation}\label{Eq:nonemit}
P_\chi(\chi_e) = e^{-\tau} \quad\quad \tau = \int \lambda_{\gamma}(\chi_e) dt. 
\end{equation}
where $\lambda_\gamma$ is the photon emission rate. In the limit $\chi_e\gg 1$, the emission rate is given by \cite{Gonoskov_RevMod_2022},
\begin{equation}
    \lambda_\gamma(\chi_e) = \frac{1.46\alpha c\chi_e^{2/3}}{\lambdabar_c\gamma},
\end{equation}
where $\lambdabar_c$ is the reduced electron Compton wavelength. If we consider the head-on collision of an electron with a circularly polarized pulse for simplicity ($a$ does not vary with the laser cycle), the laser normalized vector potential will have the form,
\begin{equation}
    a(\phi) = a_0\exp{[-\ln(2)\phi^2/(2\pi^2N^2)]},
\end{equation}
where $\phi = 2\omega_0t$ is the phase and $N=\omega_0\tau_L/2\pi$ is the number of cycles within a pulse with a full-width half-maximum (FWHM) duration $\tau_L$. Transforming Eq.~\eqref{Eq:nonemit} to an integral over phase and integrating from $-\infty$ to 0 results in,
\begin{equation}\label{Eqn:P_ne}
    \tau = 1.46\alpha \left(\frac{a_0 c}{4}\sqrt{\frac{m}{\gamma\omega_0\hbar}} \right)^{2/3} \sqrt{\frac{3\pi^3N^2}{\ln(2)}}.
\end{equation}
Now note that $\tau \propto a_0^{2/3}N\gamma^{-1/3}\omega_0^{-1/3}$, therefore $\tau^X/\tau^O = \xi^{2/3}\Omega^{-1/3}N^X/N^O$. This gives the following result,
\begin{equation}
    P_\chi^X = (P_\chi^O)^{\xi^{2/3}\Omega^{-1/3}N^X/N^O}.
\end{equation}

\subsubsection{Probability of emission near the peak of the pulse}

Perturbative SFQED predicts that the highest-energy photons will be generated in the regions of highest $\chi_e$, and therefore the signatures of high-$\chi_e$ effects should be encoded in the high-energy photons. For these effects to be measurable, the electrons that make it to high $\chi_e$ must radiate near the peak, and these photons must make it to the detector without creating pairs. For the formation of photons we can simply use Eq.~\eqref{Eq:nonemit}, integrating from the peak of the pulse to the phase where the amplitude falls to $a_0/2$ and use $P_{emit} = 1 - \exp(-\tau_{emit})$. In this case,
\begin{equation}
    \tau_{emit} = \rm{erf}\left(\sqrt{\frac{2\ln(2)}{3}}\right)\tau,
\end{equation}
where erf is the error function. Notice that $\tau_{emit}$ scales the same as $\tau$ with $\xi$ and $\Omega$. 

\subsubsection{Probability of photons propagating out of the pulse}

To measure the scattered photons they must make it through the remaining part of the pulse without creating a pair through the nonlinear Breit--Wheeler process. Conveniently, in the limit of $\chi_\gamma \gg 1$ the pair creation rate only differs from the photon emission rate by a factor of approximately 4 \cite{Gonoskov_RevMod_2022}. However, $\gamma$ and $\chi$ are that of the photon, not the electron, i.e., $\gamma = \hbar\omega/(mc^2)$ and $\chi_\gamma = 2a_0\hbar^2\omega\omega_0/m^2c^4$. Note that $\omega$ is the frequency of the scattered photon and $\omega_0$ is the frequency of the laser photons.  Integrating from the peak of the pulse to infinity results in the following:
\begin{equation}\label{Eq:tau_bw}
    \tau_{BW} = \frac{0.38\alpha_f}{4} \left(\frac{2a_0c^2m}{\hbar \sqrt{\omega\omega_0}} \right)^{2/3} \sqrt{\frac{3\pi^3N^2}{\ln(2)}}.
\end{equation}
We again care about the probability of not creating pairs, which is given by $P_{BW} = \exp(-\tau_{BW})$. Note that $\tau_{BW} \propto \omega_0^{-1/3}$, therefore, compared to the optical pulse, the XUV pulse will also have a lower probability of generating pairs. 

We want both a large number of photons produced at high-$\chi_e$ to make it to the detector, and the $\chi_e$ at which the electrons radiate to be close to the value needed to test the Ritus--Narozhny conjecture. To find the region of the $a_0$, $\gamma$, and $\Omega$, parameter space that allows for these properties, we can optimize the total probability. To do this we must also give a value for the scattered photon frequency as a function of $\gamma$ and $a_0$. Conservation of energy for nonlinear Compton scattering shows that photons scattered along the electron propagation direction have the highest frequencies, given by
\begin{equation}\label{Eq:scattered_freq}
    \omega = \frac{s\omega_0\gamma(1 + \beta)}{\gamma(1 - \beta) + 2(\frac{s\hbar\omega_0}{mc^2} + \frac{a_0^2}{2\gamma(1 + \beta)})},
\end{equation}
where $s$ is the number of scattered photons. Seipt \textit{et al.} showed that the average value of $s$ is given by \cite{Seipt_PRL_2017},
\begin{equation}\label{Eq:avg_s}
    \langle s \rangle = 0.54\frac{a_0^3}{1 + 1.49\chi_e^{0.59}}. 
\end{equation}
The total probability can then be found by taking the product of the three probabilities. 

From Fig.~\ref{fig:EPOCH_photon_chi}(d), Eq.~\eqref{Eq:tau_bw} fits well for the high-energy photons where the $\chi_\gamma \gg 1$ approximation is valid. For regions where $\chi_\gamma \ll 1$, the Laplace integration method can be used to integrate the low-$\chi_\gamma$ pair creation rate,
\begin{equation}
    \lambda_{BW,LC}(\chi_\gamma) = \frac{0.23\alpha_fc}{\gamma\lambdabar}\chi_\gamma\exp\left(\frac{-8}{3\chi_\gamma}\right),
\end{equation}
to give,
\begin{multline}\label{Eq:BW_low_chi}
    \tau_{BW,LC} = 0.23\alpha_f\chi_{\gamma,0}\exp\left(\frac{-8}{3\chi_{\gamma,0}}\right) \\ \times \sqrt{\frac{\pi^3N^2m^6c^{12}}{4\ln(2)a_0\hbar^6\omega^3\omega_0^3} \left(\frac{3\chi_{\gamma,0}^2}{3\chi_{\gamma,0} + 8}\right)   }.
\end{multline}
In Fig.~\ref{fig:EPOCH_photon_chi}(d), $100(1 - \exp(-\tau_{BW,LC}))$ is plotted, i.e., the percentage of photons that generate pairs.

\subsection{OSIRIS Photon Acceleration Simulation}

To illustrate the expected frequency upshifts and $a_0$ values that could be produced by a photon accelerator, we conducted quasi-3D, boosted-frame PIC simulations using \textsc{Osiris} \cite{OsirisRef,Davidson_JCP_2015,Yu2016}, similar to those presented in Miller \textit{et al.}, \cite{Miller_inprep}. A 50-GeV, 5.8-nC electron beam with longitudinal width 0.64~$\mu$m, transverse spot size 13~$\mu$m, and normalized emittance 1.2~mm-rad drove a wake in a 2-cm tailored density down-ramp~\cite{Sandberg_PRL_2023} beginning at $n_0 = 7\times 10^{19}$~cm$^{-3}$. A 1.2-J, 800-nm, circularly polarized witness pulse with duration 31~fs, spot size 5.5~$\mu$m, and peak $a_0 = 2.8 $ is shown in Fig.~\ref{fig:photon_accel}(a) to be upshifted by a factor of 22 
to an intensity of $6.7 \times 10^{20}$~W/cm$^2$. 
This pulse was then modeled to be focused by a plasma lens \cite{Palastro_PoP_2015,Li_PRR_2024} at $f$/3 with a focal length of 38~$\mu$m, accomplished with a Fresnel integral into the far field followed by propagation with the unidirectional pulse propagation equation \cite{Kolesik2002}. The focused pulse achieved a peak $a_0$ of 8, as seen in Fig.~\ref{fig:photon_accel}(b).  The PIC simulations used a boost of $\gamma_\mathrm{b} = 15$, two cylindrical modes (modes 0 and 1), a dispersionless field solver that also suppresses numerical Cherenkov radiation~\cite{Li2021}, a box size of $719 \times 120\,c/\omega_\mathrm{p}$ with $14896 \times 1200$ cells, 16 and 8 particles per cell for electrons and ions, respectively, and a time step of $0.027\,\omega_\mathrm{p}$, where $\omega_\mathrm{p}$ is normalized to $n_0$.
\newline

\subsection{EPOCH Colliding Beam Simulations}

To validate the theoretical results, simulations were run in the PIC code \textsc{Epoch} \cite{arber.ppcf.2015}. This code was chosen because it includes the lowest-order SFQED effects, nonlinear Compton scattering and nonlinear Breit--Wheeler pair creation. These processes are calculated using the locally constant crossed-field approximation (LCFA) \cite{reiss.jmp.1962,ritus.jslr.1985,baier.book.1998} and are implemented up to very large $\chi$. The simulations were run in 1D, allowing for large parameter space scans, with a circularly polarized laser pulse injected from one boundary and an electron beam propagating from the other side of the box. A 70-$\mu$m box was used with $\sim 46$ cells per laser wavelength. The electron beam was initialized with an energy of 50 GeV and $10^5$ particles. The electrons were allowed to propagate through the laser pulse and generate photons. We modified \textsc{Epoch} to store the maximum $\chi_e$ of each electron, as well as the $\chi_e$ of electrons and positrons when they generate photons. The photon $\chi_e$ is passed to the pairs when they are generated.

\section*{Data Availability}

The datasets generated in this study can be obtained from the corresponding author upon reasonable request.

\section*{Code Availability}

All codes used in this study are available from the corresponding author upon reasonable request.

\section*{Acknowledgements}

This work was supported by the National Science Foundation and Czech Science Foundation under NSF-GACR collaborative grant 2206059 from the NSF and Czech Science Foundation Grant No. 22-42963L. Additionally supported by NSF grant 2108075. B.K.R received additional support from US Department of Energy High-Energy-Density Laboratory Plasma Science program under Grant No. DE-SC0020103. C.P.R. was supported by UK EPSRC grant number EP/V049461/1 and funding from ELI-ERIC. S.S.B. was supported by U.S. Department of Energy Office of Science Office of High Energy Physics under Contract No. DE-AC02-05CH11231. K.G.M. was supported by the Office of Fusion Energy Sciences under Award Number DE-SC0021057. Simulations were performed at NERSC under m4372.

\section*{Author contributions}

A.G.R.T. and C.P.R. conceived the idea for the project. B.K.R. and S.S.B. led the theoretical analysis with support from A.G.R.T. and C.P.R.. B.K.R. performed the \textsc{Epoch} simulations with support from Q.Q. and analyzed and visualized the resulting data. C.A. and B.K.R. modified the \textsc{Epoch} code. K.G.M. performed the \textsc{Osiris} simulation and analyzed the resulting data with J.P.P, who also designed the plasma lens for focusing of the simulated XUV pulse. B.K.R., S.S.B, and A.G.R.T were the principal writers. All authors contributed significantly to the ideas presented in the manuscript through discussions during the project and review of the manuscript.

\section*{Competing Interests}

The authors declare no competing interests.

\bibliography{XUV.bib}

\end{document}